*Full Paper*

# Fuzzy Logic Control Based QoS Management in Wireless Sensor/Actuator Networks

Feng Xia [1,3], Wenhong Zhao [2], Youxian Sun [3] and Yu-Chu Tian [1,*]

1 Faculty of Information Technology, Queensland University of Technology,
  Brisbane QLD 4001, Australia; E-mail: f.xia@ieee.org; y.tian@qut.edu.au
2 Precision Engineering Laboratory, Zhejiang University of Technology, Hangzhou 310014, China
3 State Key Laboratory of Industrial Control Technology, Zhejiang University,
  Hangzhou 310027, China

* Author to whom correspondence should be addressed.

**Abstract:** Wireless sensor/actuator networks (WSANs) are emerging rapidly as a new generation of sensor networks. Despite intensive research in wireless sensor networks (WSNs), limited work has been found in the open literature in the field of WSANs. In particular, quality-of-service (QoS) management in WSANs remains an important issue yet to be investigated. As an attempt in this direction, this paper develops a fuzzy logic control based QoS management (FLC-QM) scheme for WSANs with constrained resources and in dynamic and unpredictable environments. Taking advantage of the feedback control technology, this scheme deals with the impact of unpredictable changes in traffic load on the QoS of WSANs. It utilizes a fuzzy logic controller inside each source sensor node to adapt sampling period to the deadline miss ratio associated with data transmission from the sensor to the actuator. The deadline miss ratio is maintained at a pre-determined desired level so that the required QoS can be achieved. The FLC-QM has the advantages of generality, scalability, and simplicity. Simulation results show that the FLC-QM can provide WSANs with QoS support.

## 1. Introduction

In the last decade, wireless sensor networks (WSNs) have been growing rapidly in various applications. Significant effort has been made in both academia and industry to meet the vision of a sensor-rich world [1-4]. Wireless sensor nodes equipped with sensing, computing, and communication capacities are now available. Typical examples include UC Berkeley's Telos and Mica family, CMU's FireFly, Intel's IMote2, Sun's SPOT, UCLA's Medusa, and MIT's µAMPS. Commercial sensor node products and solutions are also offered by many vendors, e.g., Crossbow, Rockwell, MicroStrain, Ember, Sentilla, and Dust Networks. While their physical sizes continue to decrease, these sensor node products are becoming cheaper and more powerful than ever. The availability of these products makes it possible to deploy WSNs at a large scale and a low cost that were impractical or even unimaginable just a few years ago.

WSNs are typically used for information gathering in applications like habitat monitoring, military surveillance, agriculture and environmental sensing, and health monitoring. The primary functionality of a WSN is to sense and monitor the state of the physical world. In most cases, they are unable to affect the physical environment. However, in many applications, observing the state of the physical system is not sufficient, it is also expected to respond to the sensed events/data by performing corresponding actions on the system. This stimulates the emergence of wireless sensor/actuator networks (WSANs) [5,6]. Featuring coexistence of sensors and actuators, WSANs enable the application systems to sense, interact, and change the physical world. They can be deployed in lots of applications such as disaster relief, planet exploration, intelligent building, home automation, industrial control, smart spaces, pervasive computing systems, and cyber-physical systems.

Real-world WSAN applications have their requirements on the quality of service (QoS). For instance, in a fire handling system built upon a WSAN, sensors need to report the occurrence of a fire to actuators in a timely and reliable fashion; then, the actuators equipped with water sprinklers will react by a certain deadline so that the situation will not become uncontrollable. Both delay in transmitting data from sensors to actuators and packet loss occurring during the course of transmission may potentially deteriorate control performance of the system, and may not be allowed in some situations where the systems are safety-critical. In a smart home, although there is no hard real-time constraint, actuators should turn on the lights in a timely fashion once receiving a report from sensors when someone enters or will enter a room where all lights are off; people would get unsatisfied if kept staying in dark for a long time waiting for lighting. In practice, QoS requirements differ from one application to another; however, they can be specified in terms of reliability, timeliness, robustness, trustworthiness, and adaptability, among others. Some QoS metrics may be used to measure the degree of satisfaction of these services. Technically, QoS can usually be characterized by, e.g., delay and jitter, packet loss, deadline miss ratio, and/or network utilization (or throughput) in the context of WSANs.

Meeting QoS requirements in WSANs is difficult [2,7]. Some major challenges are described as follows.

1) WSANs are normally resource constrained. Sensor nodes are usually low-cost, low-power, small devices equipped with limited data processing capability, transmission rate, energy, and memory. Due to the limitation in transmission power, the available bandwidth and the radio range of the wireless channel are also limited. For instance, the MICAz mote from Crossbow, one of the most widely-used sensor nodes, supports a data rate up to 250 kbps, which is among the highest data rates available today. However, this is far lower than the data rate offered by WLAN (up to 11 Mbps for IEEE 802.11b and up to 54 Mbps for 802.11g), and even Bluetooth (up to 3 Mbps for Bluetooth 2.0). While actuator nodes typically have stronger computation and communication capabilities and more energy budget relative to sensors, resource constraints apply to both sensors and actuators.
2) WSANs are highly dynamic in nature. The network topology may possibly change over time due to node mobility, node failure, node addition, and exhausted battery energy. The channel capacity may also change because of the dynamic adjustment of transmission powers of the sensor/actuator nodes.
3) WSANs feature inherent node heterogeneity. Having different functionality, sensors and actuators do not share the same level of resource constraints. The coexistence of sensors and actuators makes WSANs and WSNs fundamentally distinct.
4) WSANs typically operate in unpredictable environments. With wireless radio as the medium for data transmission, most WSANs suffer from diverse radio interferences. This problem will become increasingly severer as wireless technologies are incorporated in more and more (consumer) products that are expected to become pervasive. Furthermore, query-driven and event-driven applications can also cause the traffic load on the network to vary unpredictably.

This paper deals with QoS management in WSANs. A fuzzy logic control based QoS management (FLC-QM) paradigm will be developed to facilitate QoS support in resource-constrained WSANs operating in dynamic and unpredictable environments. This approach is by no means an almighty solution to all of the above challenges; it is, however, the first attempt to explicitly address the impact of unpredictable variations in traffic load on the QoS of WSANs. The variability of traffic loads over wireless connections may be a natural result of network topology changes, ambient interferences, and/or system reconfiguration, just to mention a few. The deadline miss ratio for data transmission is used as a metric to measure the QoS of WSAN. A fuzzy logic controller is designed to dynamically adjust the sampling period of relevant sensor in a way that the deadline miss ratio is kept at a desired level. Taking advantage of the feedback control technology, the FLC-QM can provide QoS guarantees while achieving predictable application performance. This solution is generic, scalable, and easy to implement. It can simultaneously address multiple QoS problems such as delay, packet loss, and network utilization. Simulation results will be given to demonstrate the effectiveness of the proposed FLC-QM scheme.

The rest of this paper is organized as follows. Section 2 reviews some related work. The architecture of the FLC-QM scheme is described in Section 3. In Section 4, the fuzzy logic controller is designed. Comparative simulations are conducted in Section 5. Section 6 concludes the paper.

## 2. Related Work

Regardless of great progress in WSN research and development, limited work has been found in the open literature on WSANs. Some QoS issues in WSNs have been addressed in e.g. [2,8], but QoS management in WSANs remains an important issue yet to be explored. Ngai *et al* [9] suggested a real-time communication framework to support event detection, reporting, and actuator coordination in WSANs. The framework takes into account the heterogeneous characteristics and functionalities of sensors and actuators. Boukerche *et al* [10] presented a QoS-aware routing protocol with service differentiation for WSANs. Morita *et al* [11] developed a reliable data transmission protocol for lossy and resource-constrained WSANs. Gungor *et al* [12] studied the impact of several network parameters on overall network performance via simulations. Zhou *et al* [13] presented a power-controlled real-time data transport protocol for energy-efficient and real-time transmission of packets. Wark *et al* [14] deployed a real-world mobile WSAN for animal control in cattle breeding industry. The mobile WSAN is capable of estimating the dynamic states of bulls, and performing real-time actuation on the bulls from location and velocity observations. Trustworthiness issue in WSANs has been discussed in [15]. However, the QoS management issue has not been addressed in any of these works in terms of deadline miss ratio and/or network utilization.

In our previous work [5], an application-level design methodology was proposed for WSANs in mobile control applications. In [16], a flexible time-triggered sampling scheme was also developed for wireless control systems. However, none of our previous reports have exploited fuzzy logic control based approach.

Another area closely related to this work is the application of fuzzy logic control to resource management in real-time computing and communication systems. In the literature, the use of control-based methods for resource management is also called feedback scheduling [17-19]. Fuzzy logic control based feedback scheduling methods have been explored in our previous work [20-22] for embedded real-time control systems. In recent years, fuzzy logic control has also been widely applied in network congestion control, e.g., [23]. Diao *et al* [24] proposed an approach to automating parameter tuning in web servers using a fuzzy controller. However, these papers have not explicitly dealt with WSANs. To the best of our knowledge, this paper is the first attempt to apply fuzzy logic control to QoS management in WSANs.

## 3. QoS Management Architecture

In a WSAN, as shown in Figure 1, there are typically lots of sensors coexisting with multiple actuators. Sensors collect information about the state of physical environment, such as the temperature and light inside a room, the occurrence of a fire, and the velocity of a mobile robot, and send corresponding messages to actuators via the wireless channel. Upon receipt of the sensed data, actuators make a decision on how to react and perform the actions on the physical world accordingly. The data transmission from a sensor to an actuator can be in a single-hop or multi-hop style. A sensor that generates original measurement data characterizing the state of physical world is called a *source* (sensor) node. In a multi-hop transmission, all other sensors except for the source node are *intermediate* nodes. In practice, a source node can also serve as an intermediate node for transmitting messages from other nodes. For simplicity, it is assumed that a source node needs to send its measurements to only one specific actuator. In addition to sensors and actuators, a base station, also referred to as sink, may be used for network management and node coordination (particularly actuator-actuator coordination).

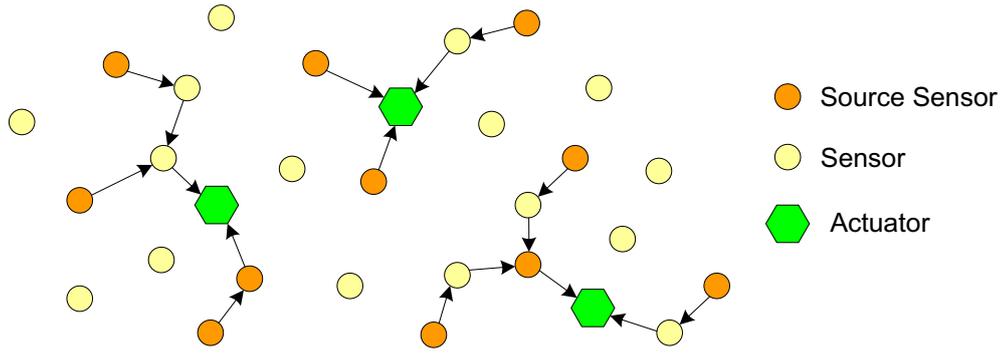

**Figure 1.** Topology of a WSAN.

The QoS of a WSAN can be affected by many factors. In the case of node movement, node removal or addition, or system update or reconfiguration, it is most likely that the network topology, routing, and node traffic load will change. This can then result in variations in network QoS attributes such as transmission delay, packet loss rate, and utilization. In some situations, the QoS of WSANs may become unsatisfactory when delay and/or packet loss rate are too large. Therefore, QoS management paradigms are needed to enhance the flexibility and adaptability of WSANs with respect to the changing network conditions.

To meet this requirement, a fuzzy logic control based QoS management (FLC-QM) scheme is proposed in this section. The basic idea of the FLC-QM scheme is to adapt the sampling period of each source sensor at run time such that the deadline miss ratio associated with the real-time data transmission from the source node to the actuator is maintained at a pre-determined desired level. Practically, both a delay larger than the deadline and a loss of packet can be regarded as deadline misses. When the sampling periods of sensors decrease, the traffic load on the network will increase. As a result, the probability of node collisions increases, leading to potential increases in both delay and packet loss rate. Therefore, increasing sampling periods can normally reduce deadline misses [16]. However, too large sampling periods will adversely cause low utilization of the network bandwidth resource. In some applications such as sampled-data control [17], smaller sampling periods may be preferable because the system performance will degrade with increasing sampling periods. For these reasons, this paper proposes to control the deadline miss ratio at a non-zero level. This can achieve high utilization of network resource while limiting the magnitudes of delay and packet loss rate within an acceptable range.

In FLC-QM, a separate QoS manager will be designed for each source sensor node to adjust its sampling period with respect to the deadline miss ratio associated with the transmission of its measurements to the actuator, as shown in Figure 2. Consider a wireless connection from source sensor $s_i$ to actuator $a_j$. There could be some or no intermediate sensors between $s_i$ and $a_j$. The QoS manager exploits the fuzzy logic control technique and operates in a time-triggered manner. Let $T_{FLC}$ denote the invocation interval of the fuzzy logic controller.

During each invocation interval, the actuator $a_j$ records the deadline misses related to data packets from $s_i$. A deadline miss occurs if $a_j$ does not receive a data packet by its deadline. At the end of each invocation interval, the deadline miss ratio DMR will be computed as:

$$DMR_i(k) = \frac{N_i(k)}{\lfloor T_{FLC}/h_i(k) \rfloor} \qquad (1)$$

where $k$ corresponds to the $k$-th invocation interval, $N_i(k)$ is the number of deadline misses recorded in this interval, $\lfloor \cdot \rfloor$ the mathematical operator rounding towards minus infinity, and $h_i$ is the sampling period of $s_i$.

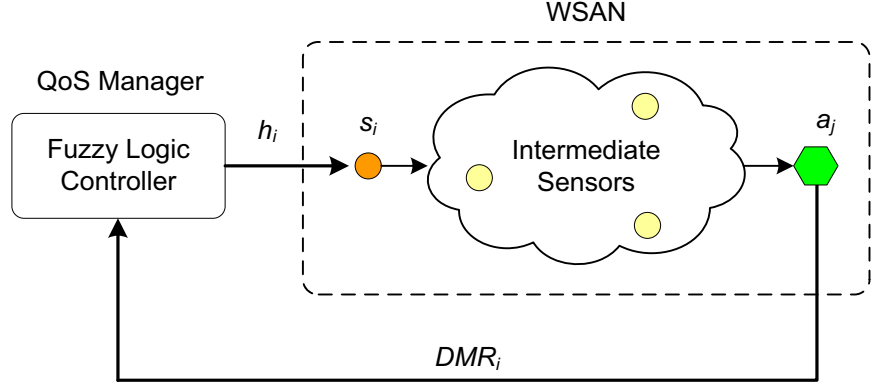

**Figure 2.** Fuzzy logic control based QoS management.

At the beginning of a new invocation interval, $a_j$ sends the value of $DMR_i(k)$ to $s_i$. With respect to this current deadline miss ratio and the desired level, the QoS manager generates the new sampling period $h_i(k+1)$ using a fuzzy logic control algorithm, which will be designed in the next section. The sampling period of $s_i$ will remain constant during the course of every invocation interval, though it might be changed at the invocation instants.

The FLC-QM scheme has the advantages of generality, scalability, and simplicity.

- *Generality*. The FLC-QM scheme is generic because it does not depend on any specific hardware (sensor nodes) or networking technologies. It is applicable to a large number of WSANs built upon different sensor/actuator nodes, with different network topologies, or using different routing and/or MAC protocols. It is well suited for various types of applications in which QoS is a concern.
- *Scalability*. The FLC-QM scheme is a distributed solution since the adjustment of sampling period is performed by a separate QoS management module for each source sensor node. When a new source node is introduced, a corresponding QoS manager can be designed for the node.
- *Simplicity*. The FLC-QM is simple because the fuzzy logic control algorithm used in the FLC-QM is computationally-cheap and is easy to implement. The small overhead makes it well-suited for resource-constrained systems like WSANs.

In addition, the use of fuzzy logic control [25] in QoS management in WSANs has the following potential advantages [17]:

- In fuzzy logic control, controllers are usually designed based on heuristic information that mainly comes from practitioners. Modelling of the process to be controlled is not required for fuzzy control system design. This is very important for complex systems such as WSANs where the relationship between system output (e.g. deadline miss ratio) and control input (e.g. sampling period) is very hard, if not impossible, to be formulated explicitly with mathematical equations.

This feature of fuzzy logic control makes it possible to fully exploit the potential of feedback control technology for QoS management in WSANs.
- As a formal methodology to emulate the intelligent decision-making process of a human expert, fuzzy logic control provides an effective and flexible way to arrive at a definite conclusion based on imprecise, noisy, or incomplete input information. Therefore, it can easily deal with various uncertainties inside WSANs, such as noise in the measurement of deadline miss ratio, unpredictable changes in traffic load and network topology.
- Fuzzy logic control is robust and adaptable since it can deliver good performance no matter whether or not the controlled process is linear. This powerful capability in handling non-linearity will reinforce good performance of QoS management in dynamic, unpredictable environments.

## 4. Fuzzy Logic Controller Design

In this section, the fuzzy logic controller in the proposed FLC-QM scheme (Figure 2) will be designed. For simplicity, the subscript $i$ in variables will be omitted wherever possible. As mentioned above, the role of the fuzzy logic controller is to determine the sampling period based on current deadline miss ratio and its setpoint. Figure 3 shows the inner structure of the fuzzy logic controller. There are two inputs, the deadline miss ratio control error $e(k)$ and the change in error $de(k) = e(k) - e(k-1)$. Let $DMR_R$ be the desired deadline miss ratio, then $e(k) = DMR_R - DMR(k)$. The output of the fuzzy logic controller is the change in sampling period $dh(k) = h(k+1) - h(k)$.

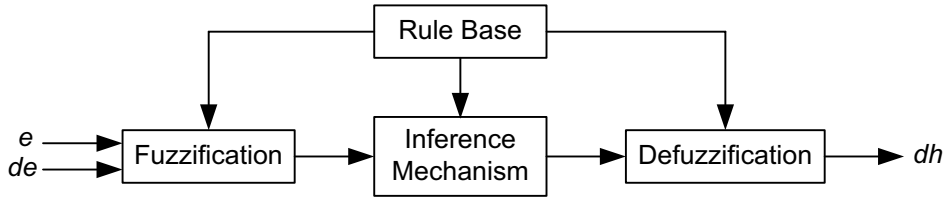

**Figure 3.** Inner structure of fuzzy logic controller.

The fuzzy logic controller is composed of four main components [25]: fuzzification interface, rule base, inference mechanism, and defuzzification interface. Once activated at the $k$-th instant, the fuzzification interface translates numeric inputs $e(k)$ and $de(k)$ into fuzzy sets characterizing linguistic variables $E$ and $DE$. The inference mechanism then applies a predetermined set of linguistic rules in the rule-base with respect to these linguistic variables, and produces the fuzzy sets of the output linguistic variable $DH$. Finally, the defuzzification interface converts the fuzzy conclusions the inference mechanism reaches to a numeric value $dh(k)$.

In this paper, the universes of discourse for $e$, $de$, and $dh$ are chosen to be [-0.2, 0.1], [-0.2, 0.2], and [-1.5, 3] (in ms), respectively. Both sets of the linguistic values for the linguistic variables $E$ and $DE$ are {NB, NS, ZE, PS, PB}, and the set of linguistic values for DH is {NB, NM, NS, ZE, PS, PM, PB}, where NB, NM, NS, ZE, PS, PM, and PB represent *negative big*, *negative medium*, *negative small*, *zero*, *positive small*, *positive medium*, and *positive big*, respectively. Figure 4 depicts the membership functions used in this paper for all linguistic values for both input and output linguistic variables. As shown in Table 1, 25 linguistic rules are built altogether.

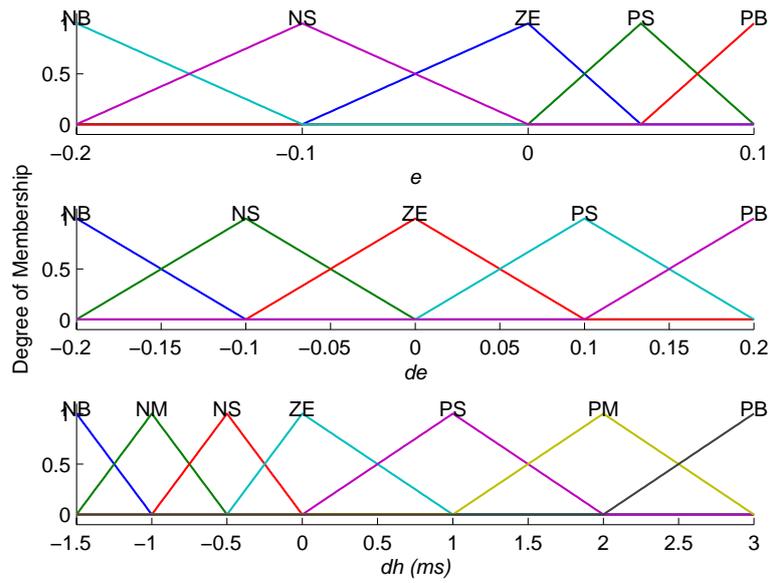

**Figure 4.** Input and output membership functions.

**Table 1.** Linguistic rules.

| DH | | DE | | | | |
|---|---|---|---|---|---|---|
| | | NB | NS | ZE | PS | PB |
| | NB | PB | PB | PB | PM | PS |
| | NS | PM | PS | PS | ZE | ZE |
| E | ZE | PS | ZE | ZE | ZE | NS |
| | PS | ZE | ZE | NS | NS | NM |
| | PB | NS | NS | NM | NB | NB |

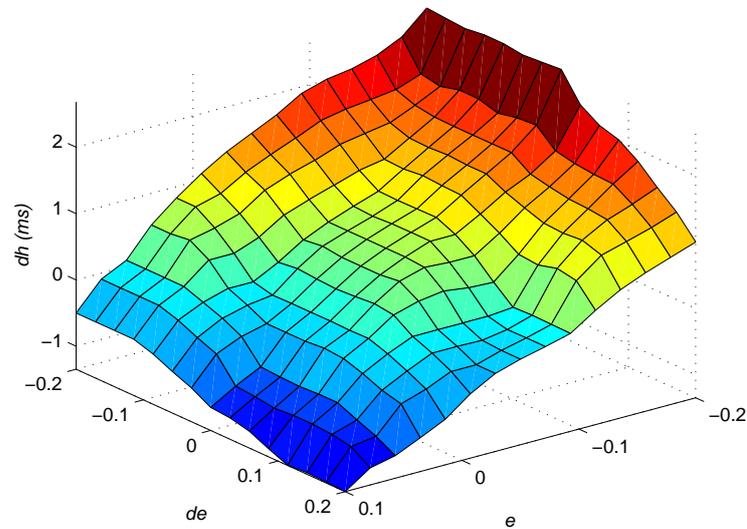

**Figure 5.** Input-output surface.

For the inference mechanism, the max-min method is adopted. In the defuzzification interface, the most popular *centre of gravity* method is used to produce a real number in the universe of discourse of the output. The input-output surface of the fuzzy logic controller is depicted in Figure 5, which describes more straightforwardly the mapping between the inputs to the output conceived by Figure 4 and Table 1.

With FLC-QM, different fuzzy logic controllers can be used in different source nodes. In particular, the deadline miss ratio setpoint and the invocation interval may be different from one another. In this way, multiple types of traffic with different QoS requirements can be supported simultaneously. For simplicity, this paper uses the same values of $DMR_R$ and $T_{FLC}$ in all QoS managers.

## 5. Performance Evaluation

Simulations are conducted in this section to evaluate the performance of the proposed FLC-QM scheme. Consider a simple yet illustrative WSAN as shown in Figure 6, where $s_1$, $s_2$, $s_3$, and $s_4$ are source sensor nodes, $s_5$ is an interfering source node, $s_6$ is an intermediate node, $a_1$ and $a_2$ are actuator nodes. These nodes reside in one collision area, that is, they have to compete for the use of the same wireless channel for data transmission. It is noteworthy that the sampling period of $s_5$ cannot be adjusted at runtime. The utilized communication protocol is ZigBee with a data rate of 250 kbps. All data packets transmitted over the network are 45 bytes in size, which may correspond to a payload of 32 bytes and an overhead of 13 bytes. The default sampling period for each source node is 10 ms, $DMR_R = 10\%$, and $T_{FLC} = 1$s. The deadline of a data packet is assumed to be equal to current sampling period of the relevant source node.

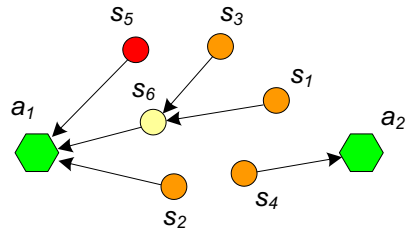

**Figure 6.** Simulated WSAN system.

The simulation runs as follows. At the beginning, all nodes except $s_3$, $s_4$, and $s_5$ are active; $s_5$ is switched on at time t = 20s and off at time t = 40s; $s_3$ and $s_4$ remains off until t = 60s. The simulation ends at time t = 80s. The simulation tool used here is Matlab along with TrueTime [26].

Figure 7 shows the deadline miss ratios corresponding to the four source nodes. With the classical design scheme, all of the deadline miss ratios are relatively high throughout the simulation. The deadline miss ratios change dramatically as the traffic over the network changes. When the interfering traffic is introduced, i.e. from t = 20s to 40s, both of the deadline miss ratios associated with $s_1$ and $s_2$ increase; particularly, the deadline miss ratio for $s_1$ reaches nearly 100% during this term. When $s_3$ and $s_4$ become active (after t = 60s), almost all messages sent by the four source nodes miss their deadlines. Further, it is found that under the same network condition the transmission from $s_1$ to $a_1$ may encounter severer deadline miss than that from $s_2$ to $a_1$. For instance, the average deadline miss ratios for $s_1$ and $s_2$ in time interval [0, 20]s are 66.5% and 37.8%, respectively. The reason behind is that the former experiences more hops than the latter. The average deadline miss ratio throughout the simulation is 81.1%, 58.4%, 100%, and 98.5%, respectively, for each source node.

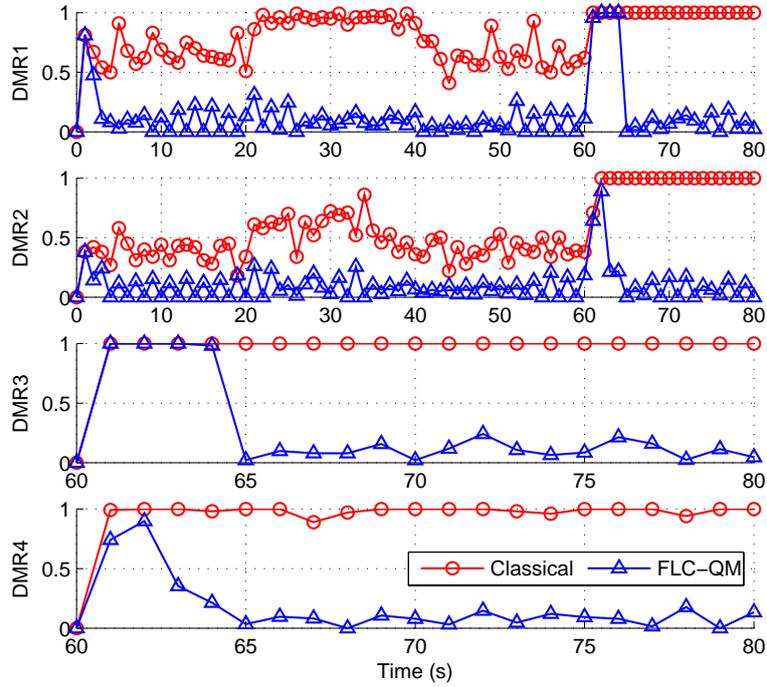

**Figure 7.** Deadline miss ratios.

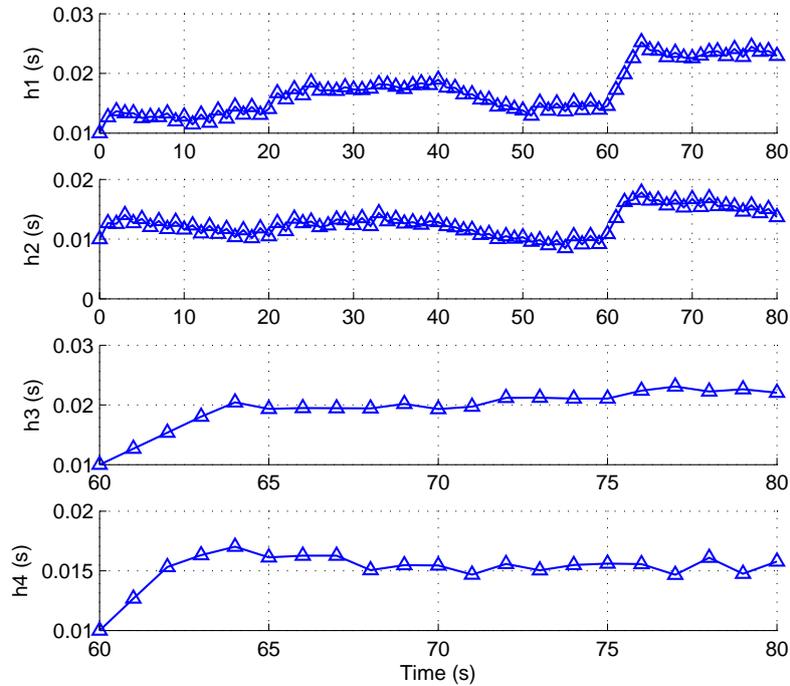

**Figure 8.** Sampling periods with FLC-QM.

When the proposed FLC-QM scheme is employed, the deadline miss ratios for all data transmissions are maintained around the desired level 10% and are much lower than those resulting from the classical scheme almost all the time (except for during the limited transient processes). The average deadline miss ratios for the four source sensors are 14.2%, 10.5%, 24.2%, and 14.2%, respectively, which are significantly lower than those associated with the classical scheme. In this case, the sampling periods of the four source sensors are adjusted dynamically at runtime, as shown in Figure 8. This is in contrast to the fixed sampling periods that used in the classical scheme and also explains why the examined two schemes perform differently in managing deadline misses.

To summarize the above simulation results, the FLC-QM scheme is effective in supporting QoS in WSANs in dynamic and unpredictable environments. It can significantly enhance the flexibility and adaptability of the systems through maintaining the desired level of QoS in terms of deadline miss ratio, and consequently delay and packet loss rate, while maximizing the network utilization as much as possible when traffic load change unpredictably.

## 6. Conclusion

A fuzzy logic control based QoS management approach has been proposed for WSANs. With this approach, the sampling period of each source sensor node is adjusted dynamically so that the deadline miss ratio associated with the relevant data transmission from the sensor to the actuator is maintained at a desired level. In this way, QoS requirements with respect to timeliness, reliability, and robustness can be satisfied. Simulation results have demonstrated the effectiveness of the proposed approach.

Our future work in this direction includes: 1) improvement of the FLC-QM scheme for large-scale WSANs through, e.g., developing a unified framework; 2) extensive simulation studies on WSANs with more complex network topology; and 3) experimental studies and practical implementation of the FLC-QM scheme in WSANs.

## Acknowledgements

Authors Xia and Tian would like to thank Australian Research Council (ARC) for its support under the Discovery Projects Grant Scheme (grant ID: DP0559111).